\documentclass[aps,prd,reprint,nofootinbib,longbibliography,preprintnumbers]{revtex4-1}

\usepackage{amsmath}
\usepackage{mathtools}
\usepackage{graphicx}
\usepackage[normalem]{ulem}
\usepackage[com pat=1.1.0]{tikz-feynman}
\usepackage{tikz}
\usepackage{aas_macros}
\usetikzlibrary{external}
\usepackage{comment}
\usetikzlibrary{arrows}
\usepackage[colorlinks=true, allcolors=blue]{hyperref}

\begin{document}
\preprint{LA-UR-24-33051}

\title{Resolution requirements for numerical modeling of neutrino quantum kinetics}

\author{Hiroki Nagakura}
\email{hiroki.nagakura@nao.ac.jp}
\affiliation{Division of Science, National Astronomical Observatory of Japan, 2-21-1 Osawa, Mitaka, Tokyo 181-8588, Japan}
\author{Masamichi Zaizen}
\affiliation{Department of Earth Science and Astronomy, The University of Tokyo, Tokyo 153-8902, Japan}
\author{Jiabao Liu}
\affiliation{Department of Physics and Applied Physics, School of Advanced Science \& Engineering, Waseda University, Tokyo 169-8555, Japan}
\author{Lucas Johns}
\affiliation{Theoretical Division, Los Alamos National Laboratory, Los Alamos, NM 87545, USA}

\begin{abstract}
Neutrino quantum kinetics is a rapidly evolving field in computational astrophysics, with a primary focus on collective neutrino oscillations in core-collapse supernovae and post-merger phases of binary neutron star mergers. In recent years, there has been considerable debate concerning resolution dependence in numerical simulations. In this paper, we conduct a comprehensive resolution study in both angular- and spatial directions by using two independent schemes of quantum kinetic neutrino transport: finite volume and pseudospectral methods. We complement our discussion by linear stability analysis including inhomogeneous modes. Our result suggests that decreasing spatial resolutions underestimates the growth of flavor instability, and then leads to wrong asymptotic states of flavor conversions, which potentially has a critical impact on astrophysical consequences. We further delve into numerical results of low resolution simulations, that reveals the underlying mechanism responsible for numerical artifacts caused by insufficient resolutions. This study settles the debate on requirements of resolutions and serves as a guideline for numerical modeling of quantum kinetic neutrino transport.
\end{abstract}
\maketitle

\section{Introduction}\label{sec:intro}
Neutrinos are dominant carrier of energy, momentum, and lepton number in the inner region of core-collapse supernova (CCSN) and binary neutron star merger (BNSM). They play a crucial role in the evolution of the system, influencing fluid dynamics and nucleosynthesis. As neutrinos propagate from inner to outer regions, they could undergo flavor conversions due to neutrino self-interactions. These self-induced neutrino flavor conversions, also known as collective neutrino oscillations, remain an unresolved complexity in the theory of CCSN and BNSM (see recent reviews, e.g., \cite{2021ARNPS..71..165T,2022arXiv220703561R,2024PrPNP.13704107F,2024RvMP...96b5004V,2024PJAB..100..190Y}). The governing equation which describes the interplay between neutrino transport, matter interactions, and flavor conversions is given by quantum kinetic equation (QKE). However, due to the nonlinear and complex nature of these physical processes, deriving analytic solutions is impractical, necessitating numerical simulations. Yet these first-principles approaches face a formidable challenge, as flavor conversions could occur on timescales of nanoseconds and spatial scales of subcentimeter, while the CCSN/BNSM simulations needs to span seconds and thousand kilometers. Addressing the issue is a daunting task but critical to advancing our understanding of CCSN and BNSM.

During the last few years, various approaches have been proposed and demonstrated for global simulations of CCSN and BNSM with neutrino flavor conversions. These can fall into four categories: (1) steady-state models (with symmetric conditions), (2) phenomenological (or parametric) models, (3) subgrid/coarse-grained models, and (4) direct numerical simulations. The first approach has been in use for many years (see e.g., \cite{2010ARNPS..60..569D} as a representative review), where time-independent QKE is solved with simplified setup such as neglecting collision term and assuming spherical symmetry. The so-called bulb model, where neutrinos are assumed to be radiated from a neutrino sphere and then freely propagate outward, belongs to this category. While this approach provided many insights into some global aspects of collective neutrino oscillations, it suppresses temporal flavor instabilities \cite{2015PhRvD..92l5030D}. We also note that neglecting neutrino-matter interactions and spherically symmetric conditions decrease the fidelity of neutrino transport, indicating that the obtained steady-state neutrino radiation field would be different from realistic CCSN models.

The second approach, phenomenological model, is commonly employed by current CCSN/BNSM modelers. In this approach, effects of flavor conversions are treated in a parametric way \cite{2021PhRvL.126y1101L,2022PhRvD.105h3024J,2022PhRvD.106j3003F,2023PhRvD.107j3034E,2023PhRvL.131f1401E,2024PhRvD.109j3017N}. This approach can offer qualitative understandings of how neutrino flavor conversions impact on global dynamics of CCSN and BNSM. One of the noticeable advantages of this approach is its easy implementation into classical neutrino transport. Nevertheless, there is a crucial limitation that it does not have abilities to pinpoint the precise locations, timings, and physical processes involved in flavor conversions, resulting in considerable uncertainties in this approach.

The third approach, subgrid model, has a potential to overcome the shortcoming in the phenomenological approach. Unlike phenomenological models, this approach determines occurrences and (local) asymptotic states of flavor conversions using more physically accurate techniques (see, e.g., \cite{2023PhRvD.107j3022Z,2023PhRvD.108f3003X,2023arXiv230614982J,2024PhRvD.109d3024A,2024PhRvD.110j3019R,2024arXiv240312189F}). In essence, effects of short-term and small-spatial flavor conversions are integrated into a coarse-grained neutrino transport equation in one way or another (see, e.g., \cite{2024PhRvD.109h3013N,2024arXiv240317269X}). While this approach is promising, its accuracy hinges on the fidelity of local models. One thing we do notice here is that there has been significant progress in local models such as determining asymptotic states of flavor conversions in local scales, especially for fast-neutrino flavor conversion or FFC; see, e.g.,  \cite{2018PhRvD..98j3001D,2023PhRvD.107j3022Z,2023PhRvD.107l3021Z,2024JHEP...08..225F,2024arXiv240917232F}). However, our comprehension of the underlying physical processes of collective neutrino oscillations remains incomplete \cite{2024PhRvD.109h3031Z,2024arXiv240312189F,2024arXiv241108503L}. This implies that much work is still needed to enable global simulations that adequately capture the effects of flavor conversions.

The final approach, directly solving QKE, offers a unique probe of global features of neutrino radiation field. As pointed out already, large disparities in time- and spatial scales between flavor conversions and astrophysical systems make the global numerical simulations intractable. The attenuation method addresses this challenging issue by reducing the strength of neutrino oscillation Hamiltonian, which allows us to study complex physical processes of flavor conversions in global simulation framework. Although the attenuation of Hamiltonian results in a smoothing of small scale structures of flavor conversions, the overall time-averaged profile remains relatively unaffected as long as the scale of flavor conversion is much smaller than the astrophysical one \cite{2022PhRvL.129z1101N,2023PhRvD.107f3033N}. Another key aspect of this approach is that the convergence study with respect to attenuation parameter allows us to quantify the error of the result, and the attenuated models can be extrapolated to more realistic case through the convergence study. Although these simulations still require large computational resources, there have already been multiple demonstrations for these simulations \cite{2023PhRvD.107h3016X,2023PhRvL.130u1401N,2023PhRvD.108l3003N,2024PhRvD.109l3008X}.

It should be noted, however, that other studies have argued that the attenuation of neutrino Hamiltonian is unnecessary and that standard spatial resolutions of classical neutrino transport are sufficient for global simulations (see, e.g., \cite{2023PhRvD.108d3006S,2024PhRvD.109j3011S} and references therein). This assertion is based on two main points. First, the growth rate of neutrino flavor conversions associated with flavor instabilities is good agreement in between their numerical simulations and linear stability analysis for homogeneous mode \cite{2024PhRvD.109j3011S}. Second, the neutrino oscillation pattern can be smeared out by superposition of flavor waves if neutrinos are emitted from various locations (see Fig. 7 and related discussions in \cite{2023PhRvD.108d3006S}). Consequently, the effective wavelength of flavor conversions becomes longer, while the scale is governed by the collision rate which is typically much larger than the scale of flavor conversions.

This debate, marked by conflicting views regarding requirements of spatial resolutions in numerical modeling of neutrino quantum kinetics, stems from the lack of detailed studies of flavor conversions that specifically address the resolution problem. In addition to this, it is currently unclear how low resolution simulations impact on asymptotic states of flavor conversions. This issue is, hence, worth to be investigated, which is what motivates the present study. In this paper, we present a comprehensive resolution study of FFC by following a previous study in \cite{2024arXiv240806422S}. We inspect our numerical results with the help of linear stability analysis, which exhibits mechanisms accounting for numerical artifacts due to low resolutions.

This paper is structured as follows. We begin by briefly summarizing essential parts of our numerical methods and models in Sec.~\ref{sec:methodmodel}. Before presenting the major result of this paper, we show a dispersion relation of flavor instabilities based on linear stability analysis in Sec.~\ref{sec:linearsta}, which supports our interpretation of the numerical simulations. All results in numerical simulations are detailed in Sec.~\ref{sec:results}, where we delve deeper into the results to provide insights into numerical artifacts. We summarize our conclusions in Sec.~\ref{sec:summary}. Throughout this paper, we express time- and space units in terms of $\mu$, where $\mu$ is defined as $\mu \equiv \sqrt{2} G_F n_{\nu_e}$. In the expression, $G_F$ and $n_{\nu_e}$ denote Fermi constant and number density of electron-type neutrinos ($\nu_e$), respectively. We choose the spacetime metric signature to be $(-,+,+,+)$.

\section{Method and model}\label{sec:methodmodel}
In the conventional approach for studying neutrino quantum kinetics, the time evolution and spatial variation of the neutrino radiation field are modeled by solving the mean-field QKE,
\begin{equation}
    (\partial_t+\boldsymbol{v}\cdot\nabla)\rho = - i \left[\mathcal{H},\rho\right] + C,
\label{eq:QKE}
\end{equation}
where $\rho$, $\mathcal{H}$, and $C$ denote the density matrix of neutrinos, neutrino oscillation Hamiltonian, and collision term, respectively. In this expression, $t$ and $\boldsymbol{v}$ denote time and neutrino flight direction in space, respectively. In the present study, our simulations are limited to one spatial-dimensional ($z$-direction) and monoenergetic with axisymmetric angular distributions in neutrino momentum space. We also neglect collision term and a matter potential in neutrino oscillation Hamiltonian. These simplifications help in isolating the origin of numerical artifacts that arise from insufficient resolutions.

We employ a neutrino transport code in \cite{2023PhRvD.107l3021Z}. In this code, Eq.~\ref{eq:QKE} is solved by a finite volume approach with seventh-ordered weighted essentially non-oscillatory (WENO) scheme, while the fourth-order Runge-Kutta method is used for time integration. As reported in the literature (see, e.g., \cite{2023CoPhC.28308588G}), such high-order numerical schemes are necessary to accurately capture the interplay between collective neutrino oscillations with neutrino advection.

To ensure the reliability of our numerical simulations, we utilize another QKE solver with a pseudospectral method \cite{2021PhRvD.104h3035Z}. Assuming periodic boundary conditions in space, the QKE can be written in terms of spatial Fourier components. The pseudospectral method offers exponential convergence with increasing the number of spectral modes, unless the discontinuous solution is involved. For the problem we consider in the present study, there are no discontinuities in any physical quantities, ensuring the high accuracy. We also note that a detailed code comparison among different QKE schemes has been made for FFC simulations in \cite{2022PhRvD.106d3011R}, and numerical results in the pseudospectral code showed excellent agreement with others, lending confidence to our pseudospectral scheme.

We employ the same initial condition of neutrino distributions in \cite{2024arXiv240806422S}. The angular distributions of $\nu_e$ and its antipartners ($\bar{\nu}_e$) are set as,
\begin{equation}
\begin{aligned}
&\rho_{\nu_e}(\cos \theta_{\nu}) = 0.5 , \\
&\rho_{\bar{\nu}_e}(\cos \theta_{\nu}) = 0.47 + 0.05 \exp \large( -(1 - \cos \theta_{\nu} )^2  \large),
\end{aligned}
\label{eq:iniangdistri}
\end{equation}
respectively, where $\theta_{\nu}$ specifies the neutrino flight angle with respect to $z$-direction, i.e., $v_z = \cos \theta_{\nu}$. At the beginning of simulations, the neutrino distributions are assumed to be homogeneous in space, but we add a small and random perturbation in off-diagonal components of density matrix of neutrinos. For the purpose of completeness, we also consider the case with vacuum oscillation in which all physical parameters are the same as those used in \cite{2024arXiv240806422S}. The vacuum frequency ($\omega_{\rm{vac}}$) and the mixing angle ($\theta_{\rm{mix}}$) are chosen as $\omega_{\rm{vac}} = 10^{-5} \mu$ and $\theta_{\rm{mix}} = 10^{-3}$, respectively. In all simulations, we fix the spatial domain ($L=10000 \mu^{-1}$), and the spatial resolution is controlled by the number of grid points ($N_z=10, 40, 100, 1000,$ and $10000$). According to the linear stability analysis (see in Sec.~\ref{sec:linearsta}), the model with $N_z=100$ grid point corresponds to the minimum resolution required to capture the maximum growth mode at $K = K_{\rm peak} (\sim - 0.03)$, while the model with $N_z=1000$ grid points nearly reaches the threshold to resolve all unstable modes. The simulation with $N_z=10000$ grid points can resolve the highest wave number ($K_{\rm max}$) with $\sim 10$ grid points. In the neutrino angular direction, on the other hand, the default setup is $N_v=500$ grid points. To see the dependence of angular resolution, we demonstrate two more simulations with $N_v=125$ and $250$ grid points with the highest spatial resolution ($N_z=10000$ grid points).

\section{Linear stability analysis}\label{sec:linearsta}

Given a fixed point of neutrino states, the time evolution of neutrino radiation field in the early phase can be well approximated by linearized QKE. As described in Sec.~\ref{sec:methodmodel}, all neutrinos are assumed to be nearly in flavor eigenstates at $t=0$. This condition enables us to treat the off-diagonal component of density matrix of neutrinos (and antineutrinos) as perturbed quantities. With the plane wave ansatz, the linearized QKE can be cast into the form of dispersion relation, which allows us to assess whether the system is stable or not. If there are unstable modes, the dispersion relation provides us the growth rate of flavor instabilities as a function of wave number ($K$). We refer the reader to \cite{2017PhRvL.118b1101I,2018JCAP...12..019A} for more technical details to compute the dispersion relation.

\begin{figure}
\begin{minipage}{0.5\textwidth}
    \includegraphics[width=\linewidth]{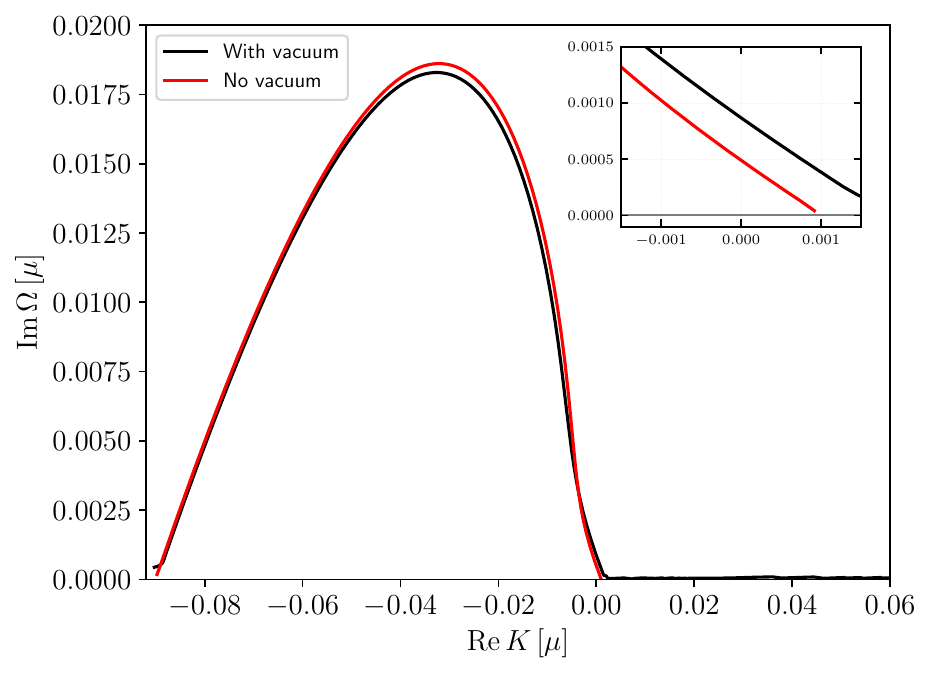}
\end{minipage}
    \caption{The growth rate (${\rm Im} \Omega$) as a function of wave number (${\rm Re} K$) for the initial angular distribution of neutrinos (see Eq.~\ref{eq:iniangdistri}). The black and red lines represent the result with and without vacuum contribution, respectively. In the top right panel, we magnify the region around ${\rm Re}K=0$.
}
    \label{DR_inset}
\end{figure}

We perform a linear stability analysis of neutrino flavor conversions for the initial condition of neutrino distributions (see Eq.~\ref{eq:iniangdistri}), where inhomogeneous modes ($K \ne 0$) are also taken into account. The dispersion relation obtained from our analysis is displayed in Fig.~\ref{DR_inset}. Two key points should be highlighted. First, homogeneous mode ($K=0$) is unstable (see the inserted panel at the upper right corner), but the growth rate is significantly lower than the maximum at $K = K_{\rm peak} (\sim - 0.03 \mu)$. Second, vacuum potentials have negligible effects for the overall trend of dispersion relation except for the region around $K=0$ where the growth rate of flavor instability becomes nearly twice larger than the case in no vacuum contributions.

The stability analysis provides a valuable insight on numerical artifacts in low resolution simulations. We shall demonstrate that the failure to resolve all unstable modes result in incorrect asymptotic states, while the property can be interpreted through linear stability analysis (see Sec.~\ref{sec:results} for more details). These detailed analyses lead us to draw a robust conclusion that the high spatial resolutions resolving all unstable modes are indispensable for accurate numerical modeling of collective neutrino oscillations.

\section{Results}\label{sec:results}
\subsection{Basic features}\label{subsec:basicfeature}

\begin{figure*}
    \begin{minipage}{0.45\textwidth}
    \includegraphics[width=\linewidth]{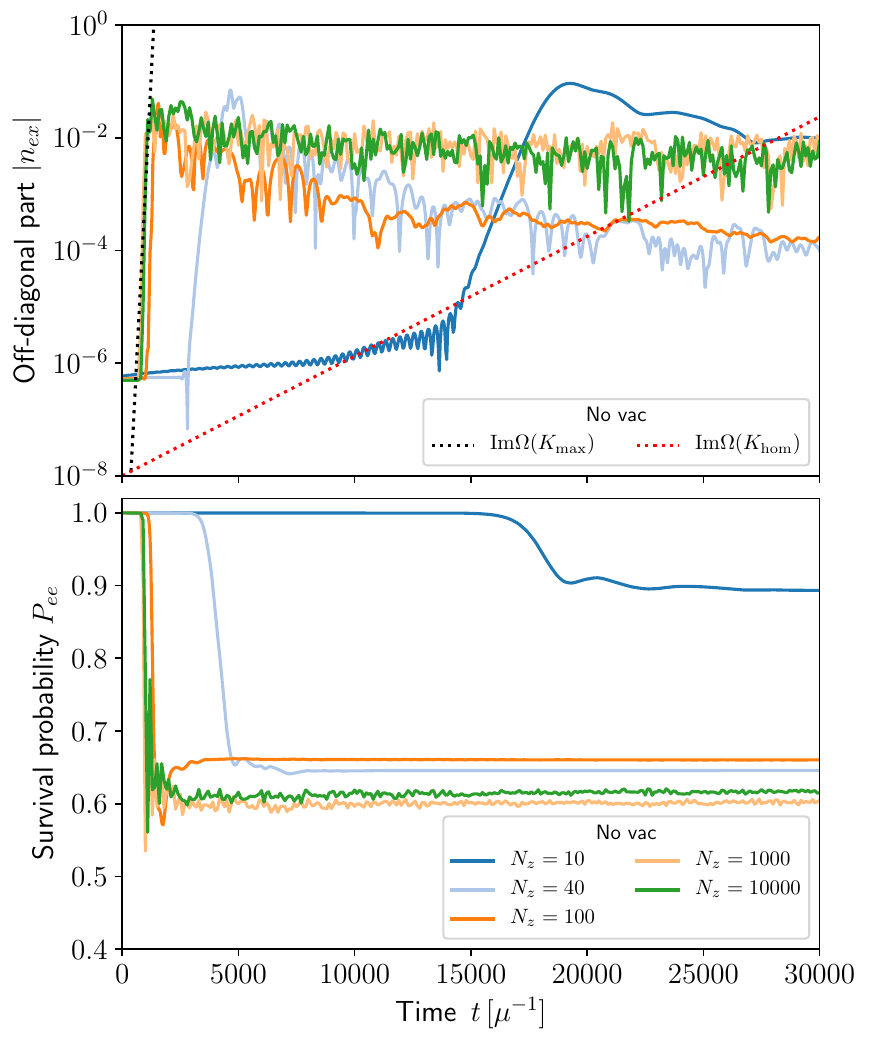}
    \end{minipage}
    \begin{minipage}{0.45\textwidth}
    \includegraphics[width=\linewidth]{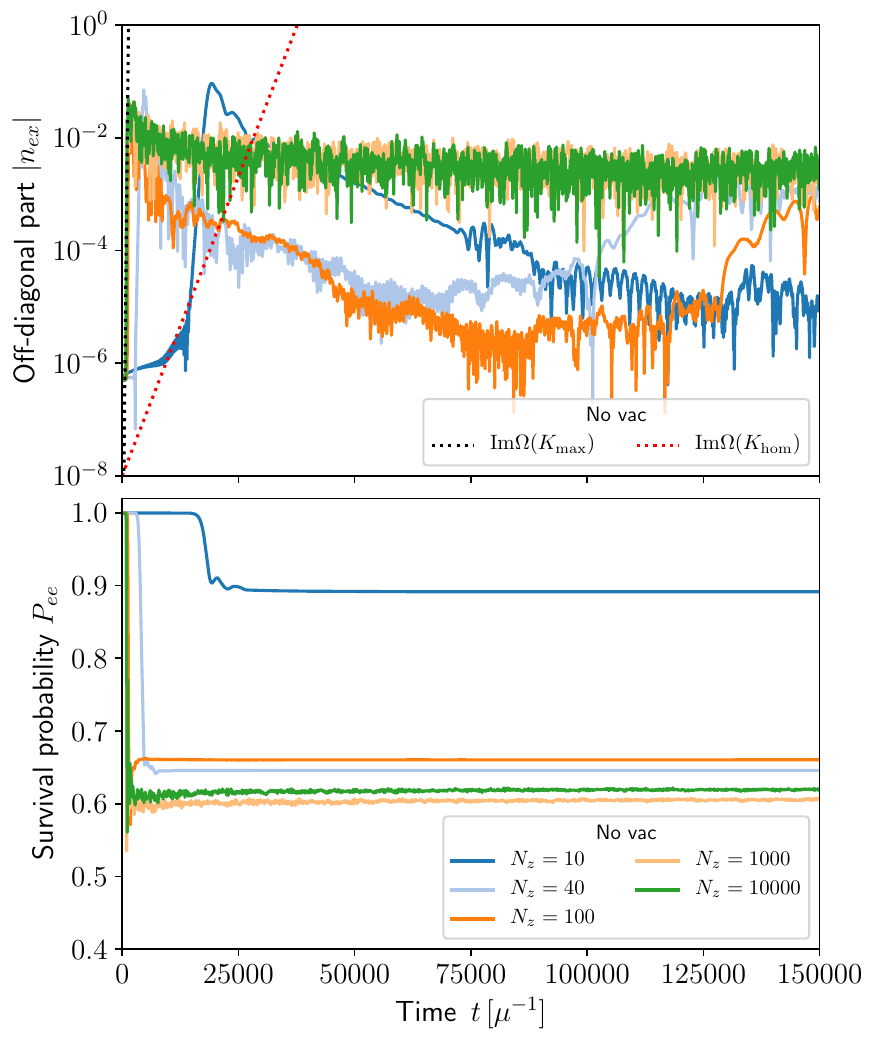}
    \end{minipage}
    \caption{Time evolution of the spatial average of flavor coherency ($|n_{ex}|$) and the survival probability of $\nu_e$ ($P_{ee}$) in top and bottom panels, respectively. The solid lines represent numerical results. In the top panels, two dashed lines represent a growth rate of flavor instability obtained from linear stability analysis. Colors in solid lines distinguish models with different spatial resolutions. The left panels focus on the early phase ($t<30000 \mu^{-1}$), while the entire evolution of these quantities are displayed in the right panels. In this plot, we only show the case without vacuum contribution.
}
    \label{evo}
\end{figure*}

\begin{figure*}
\begin{minipage}{0.45\textwidth}
    \includegraphics[width=\linewidth]{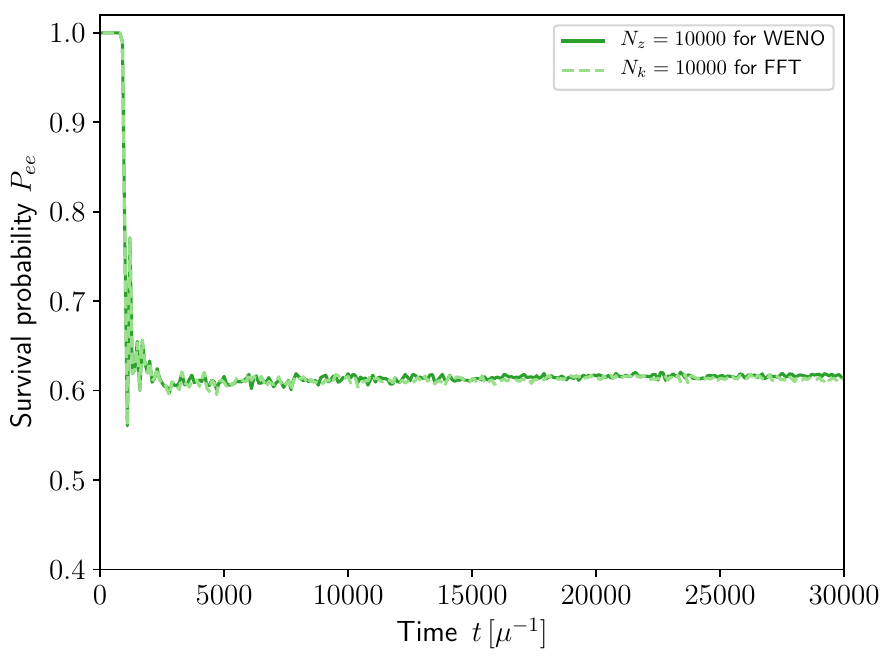}
\end{minipage}
\begin{minipage}{0.43\textwidth}
    \includegraphics[width=\linewidth]{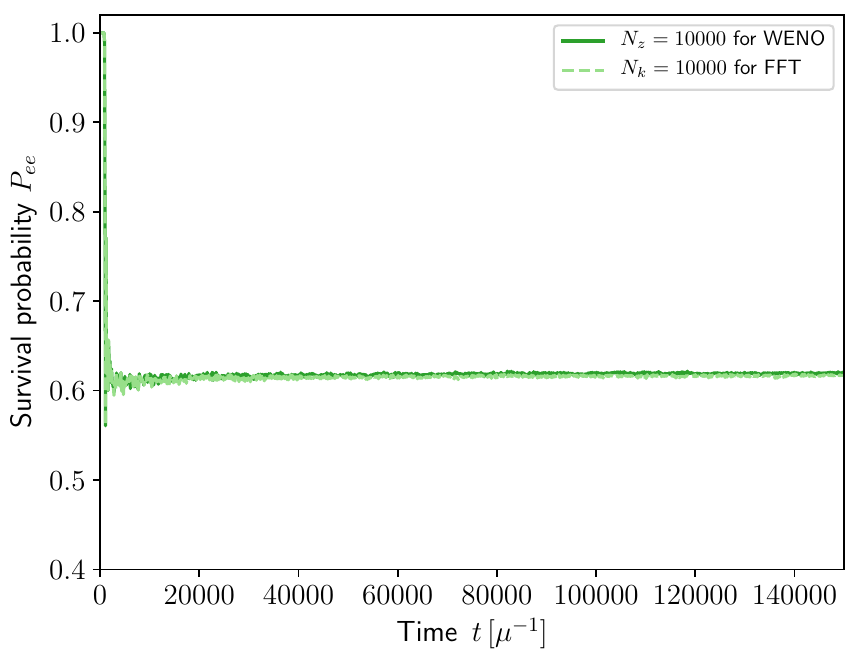}
\end{minipage}
    \caption{Same as the bottom panels in Fig.~\ref{evo} but for comparing between finite volume and pseudospectral methods. We show only the result of the highest resolution model in these plots.
}
    \label{comparison_surv}
\end{figure*}

Hereafter, we present results of our numerical simulations. In the upper panels of Fig~\ref{evo}, we display the time evolution of spatially averaged $|n_{ex}|$ for models with no vacuum contributions, where $n_{ex}$ represents the zeroth angular moment of off-diagonal component for the density matrix of neutrinos. As expected from linear stability analysis, the growth of flavor coherency in the early phase for low spatial resolution models ($N_z < 100$) is significantly underestimated (see the left panel). Consistently, in the bottom panels of Fig~\ref{evo}, which shows the time evolution of survival probability of $\nu_e$ ($P_{ee}$), we observe notable delays of onset of non-linear phase for low resolution models. This is consistent with the stability analysis in Sec.~\ref{sec:linearsta}. We also find that the asymptotic behavior strongly hinges on the spatial resolutions. Our result suggests that decreasing spatial resolutions has a devastating effect on numerical modeling of flavor conversions.

\begin{figure*}
\begin{minipage}{0.45\textwidth}
    \includegraphics[width=\linewidth]{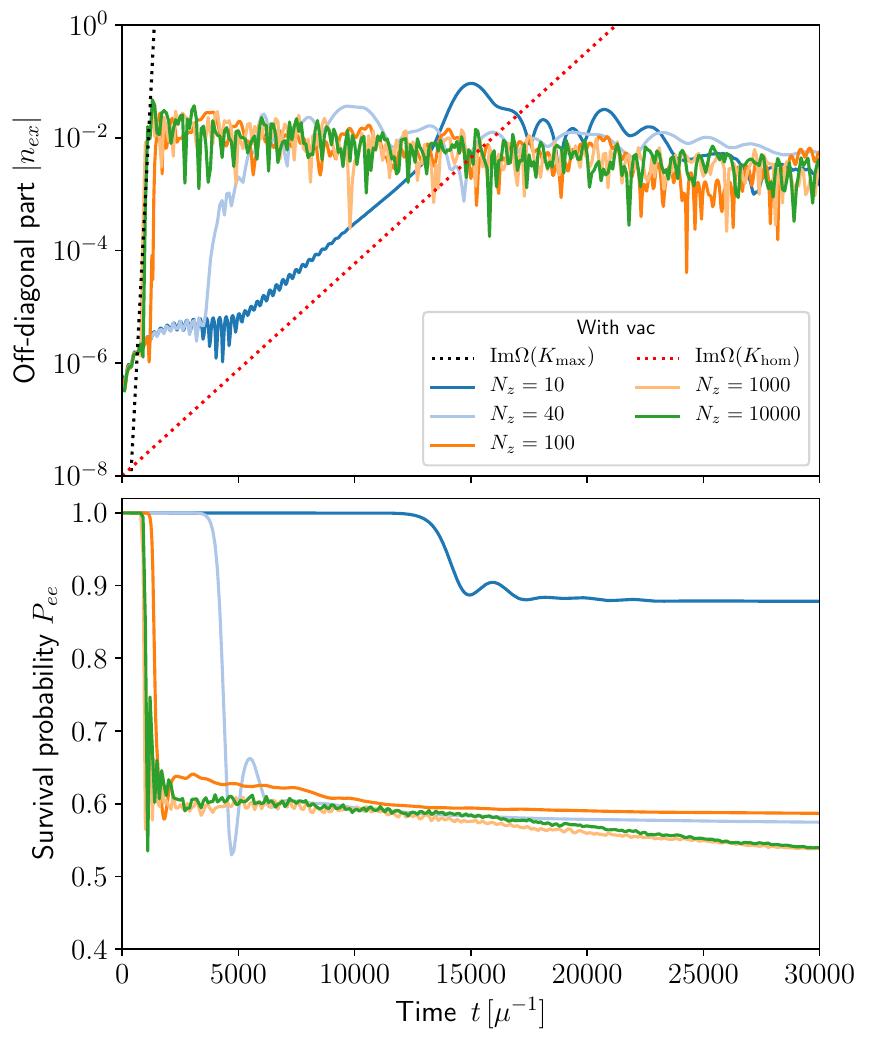}
\end{minipage}
\begin{minipage}{0.45\textwidth}
    \includegraphics[width=\linewidth]{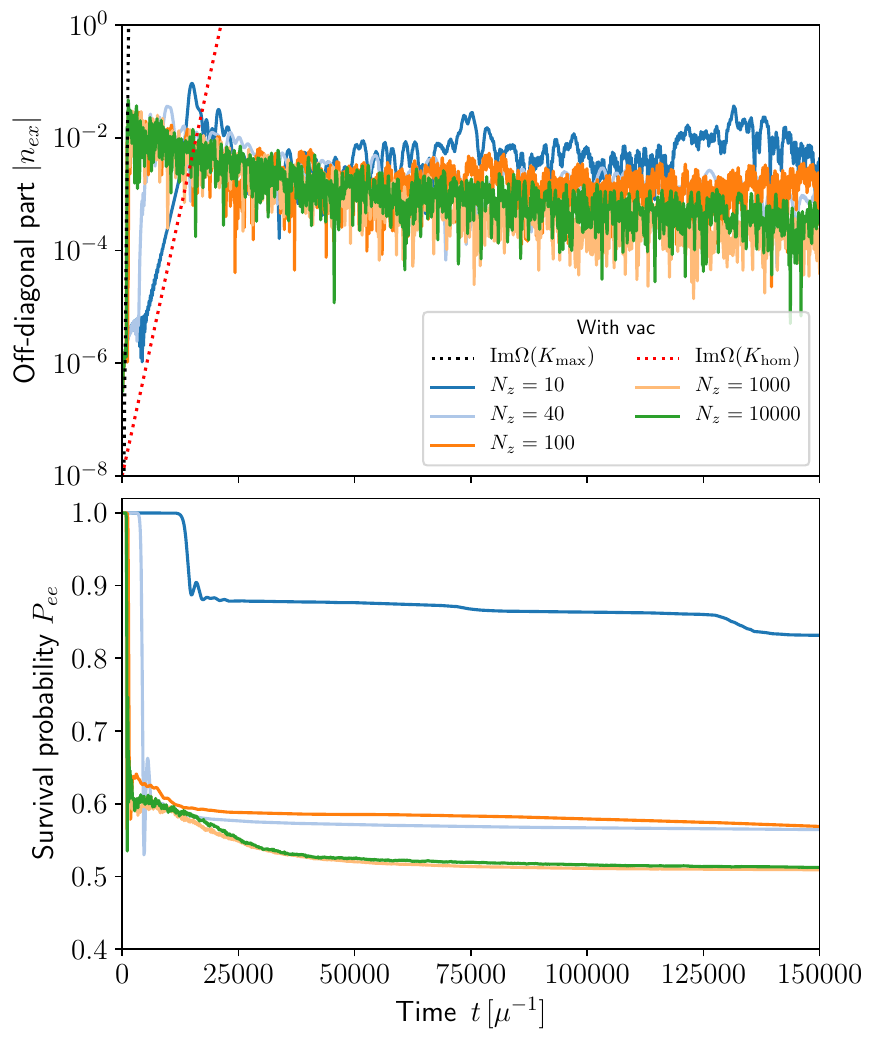}
\end{minipage}
    \caption{Same as the bottom panels in Fig.~\ref{evo} but with vacuum contribution.
}
    \label{evo_withvac}
\end{figure*}

It is worth emphasizing that the highest resolution model ($N_z=10000$) shows an excellent agreement with another simulation by pseudospectral method with $N_k=10000$, where $N_k$ denotes the number of spatial Fourier modes; see Fig.~\ref{comparison_surv}. This comparison validates the highest resolution model obtained by our finite volume scheme. We also find that the qualitative trend of resolution dependence for models with vacuum contributions is essentially the same as those without vacuum contribution (see Fig.~\ref{evo_withvac})\footnote{There is a minor difference regarding asymptotic behaviors between with/without vacuum contributions. In models that include vacuum contributions, slow modes drive flavor conversions in the late phase, after temporal asymptotic states appear by FFCs.}. In the following analyses, we, hence, focus on models with no vacuum contributions.

\begin{figure}
\begin{minipage}{0.5\textwidth}
    \includegraphics[width=\linewidth]{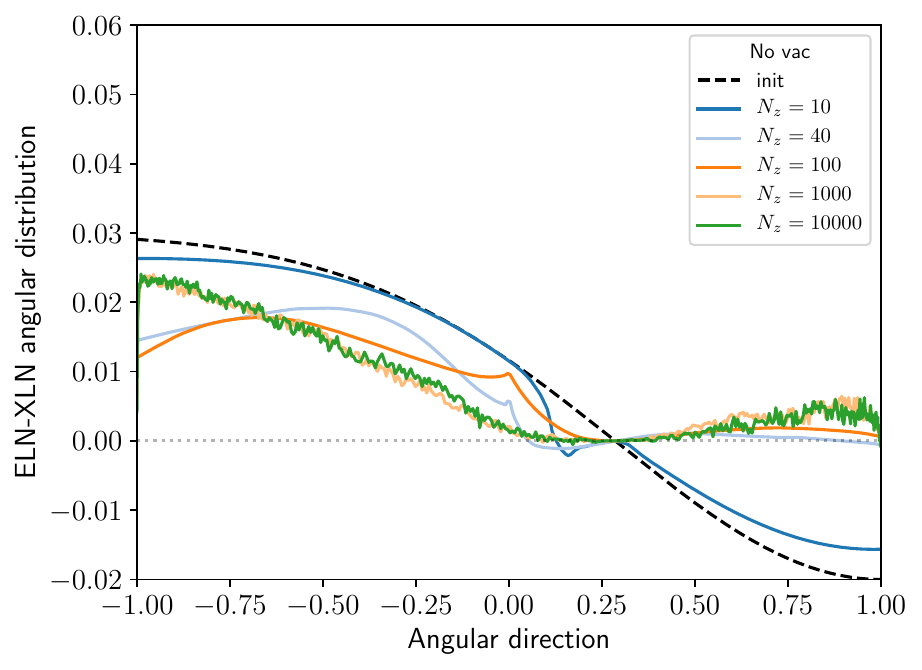}
\end{minipage}
    \caption{Spatially averaged angular distributions at the end of each simulation for the case without vacuum contribution. The color code is the same as that used in Fig.~\ref{evo}. For comparison, we also display the initial angular distribution as a dashed line.
}
    \label{ang_dist}
\end{figure}

\begin{figure}
\begin{minipage}{0.5\textwidth}
    \includegraphics[width=\linewidth]{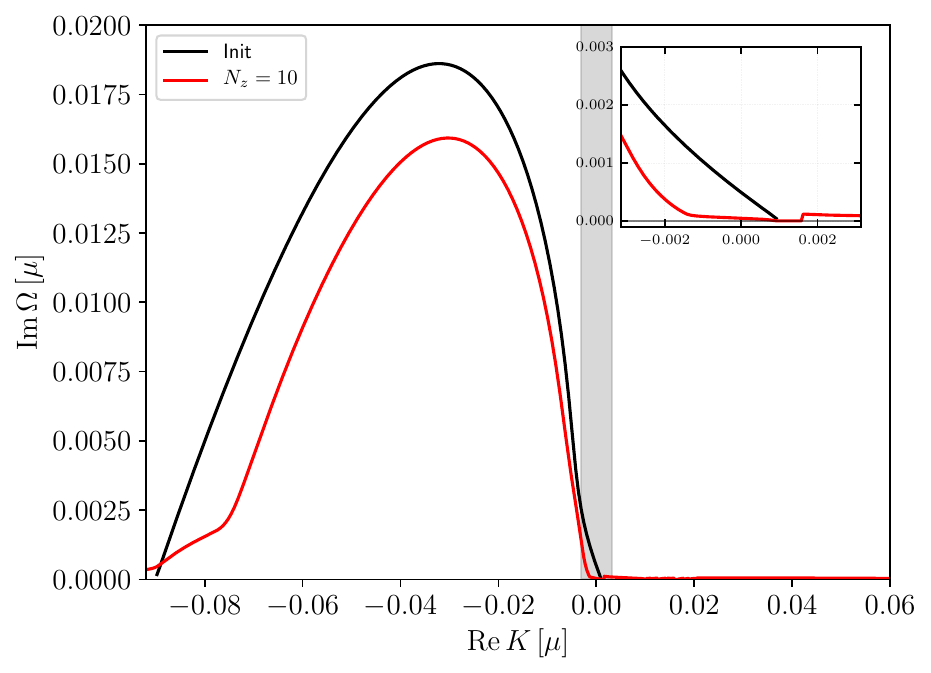}
\end{minipage}
    \caption{Same as Fig.~\ref{DR_inset} but for the model with the lowest spatial resolution ($N_z=10$). We compute the dispersion relation with respect to the asymptotic state (or at the end of our simulation) by employing the spatially averaged ELN-XLN angular distribution; the result is displayed as a red line. For comparison, we also show the dispersion relation at $t=0$ as a black line. We highlight the magnified region by shaded color in the main panel.
}
    \label{DR_novac}
\end{figure}

\begin{figure*}
\begin{minipage}{0.45\textwidth}
    \includegraphics[width=\linewidth]{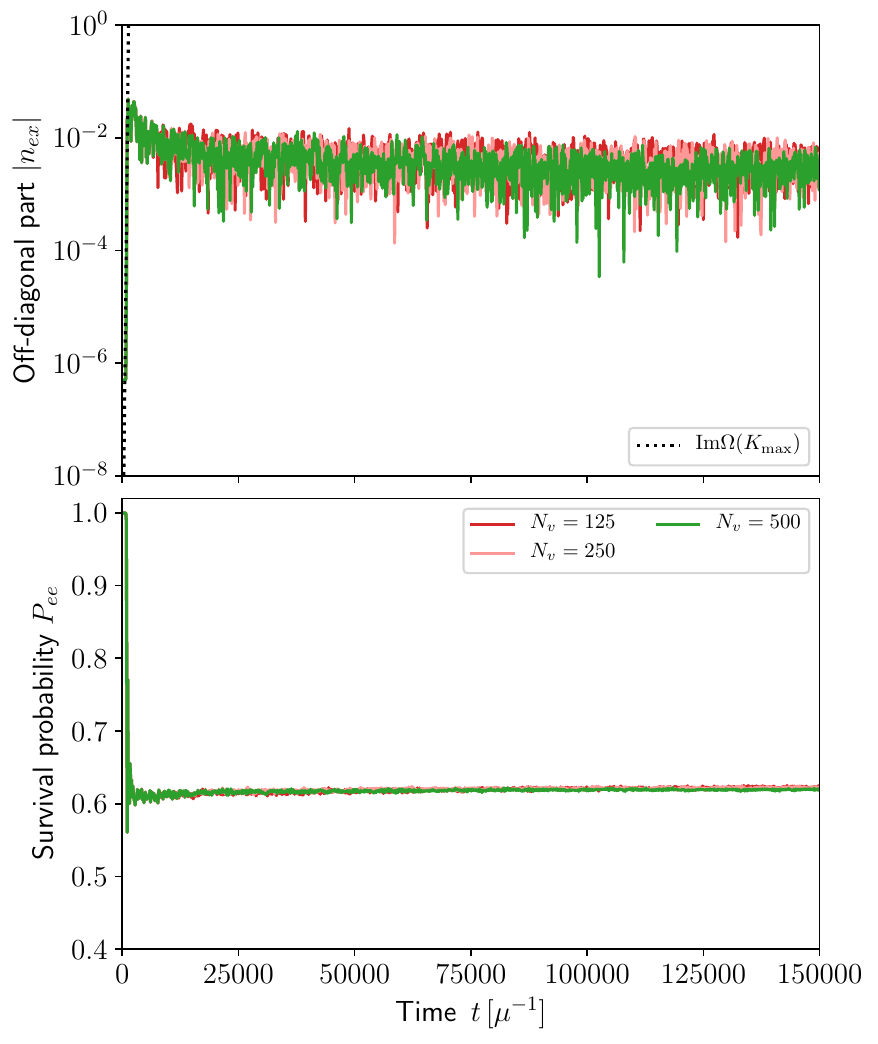}
\end{minipage}
\begin{minipage}{0.45\textwidth}
    \includegraphics[width=\linewidth]{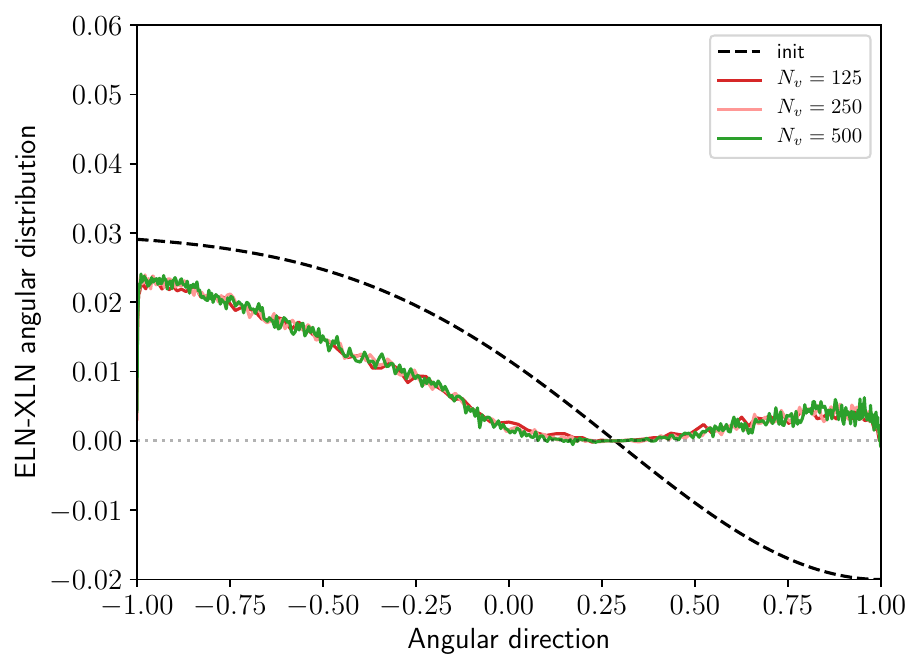}
\end{minipage}
    \caption{Similar as Figs.~\ref{evo}~and~\ref{ang_dist} but focusing on the resolution dependence in the angular direction of neutrino momentum space.
}
    \label{angdepe}
\end{figure*}

\begin{figure}
\begin{minipage}{0.5\textwidth}
    \includegraphics[width=\linewidth]{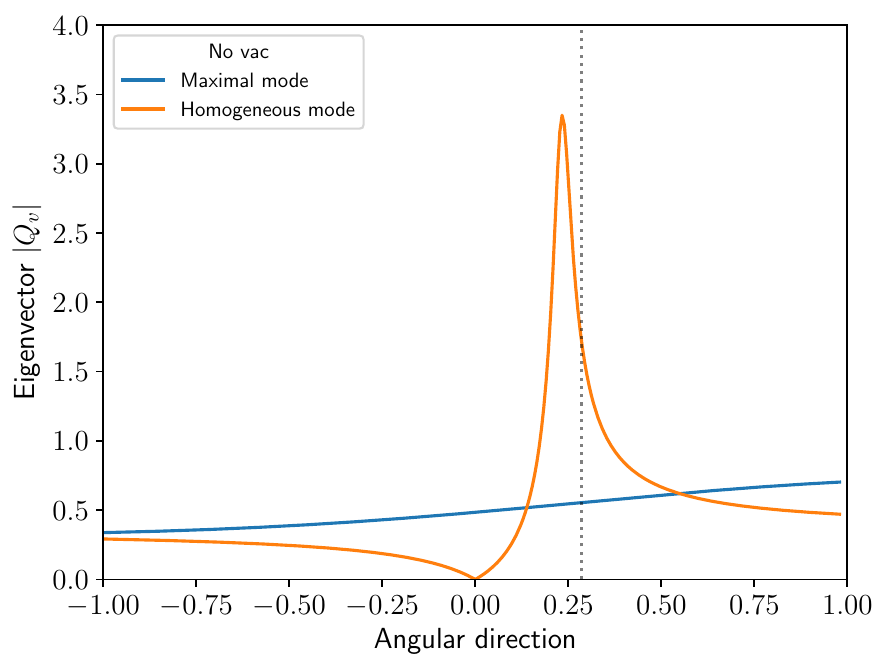}
\end{minipage}
    \caption{Angular distributions of eigenvectors, $|Q_{v}|$, for two representative unstable modes: maximal ($K=K_{\rm peak}$) and homogeneous ($K=0$) modes. The vertical scale is normalized so as to be $\int |Q_{v}| dv =1$.
}
    \label{Qv_amp_hom}
\end{figure}

It is worthwhile to inspect the neutrino angular distributions in asymptotic states among different spatial resolutions. As we shall discuss below, this analysis gives clues to understand the origin of numerical artifacts in low resolution models. In Fig.~\ref{ang_dist}, spatially averaged angular distributions of $\nu_e - \bar{\nu}_e - \nu_x + \bar{\nu}_x$, known as ELN-XLN, are displayed ($\nu_x$ denotes the heavy leptonic neutrinos). It should be mentioned that the ELN-XLN angular distribution corresponds to a key quantity to determine the asymptotic state of FFC \cite{2022PhRvL.129z1101N,2023PhRvD.107j3022Z,2023PhRvD.107l3021Z}. As shown in Fig.~\ref{ang_dist}, an ELN-XLN angular crossing, which is present at the initial distribution, disappears in high resolution models. This is consistent with other studies in the literature \cite{2023PhRvD.107j3022Z,2023PhRvD.108f3003X,2024PhRvD.110j3019R,2024arXiv240908833G}. However, in the lowest resolution model ($N_z=10$), the crossing does not fully disappear, suggesting that the strength of flavor conversions is weaker compared to other higher resolution models. This feature is also evident in Fig.~\ref{evo}. The $|n_{ex}|$ and $P_{ee}$ in the model are notably lower and higher, respectively, than those in higher resolution models.

As displayed in Fig.~\ref{ang_dist}, on the other hand, the model with $N_z=40$ is greatly improved compared to the case with $N_z=10$. In fact, the ELN-XLN angular crossing nearly disappears in the model. This improvement is linked to a sharp increase of growth rate of FFC with decreasing $K$ in the range of $K_{\rm peak} < K < 0$ (see Fig.~\ref{DR_inset}). This suggests that the increase of spatial resolutions resolves unstable modes with higher growth rates, leading to a substantial change in the time evolution of the system. It should be noted, however, that the model with $N_z=40$ still suffers from the resolution issue. There remains a clear deviation of ELN-XLN angular distribution in the angular region of $-1 < \cos \theta_{\nu} < 0$ compared to the highest resolution model. This deviation gradually diminishes with increasing spatial resolutions, and the convergence nearly achieves at $N_z=1000$. This result indicates that the accurate asymptotic state of FFC can be obtained only if all unstable modes are properly resolved.

The incomplete disappearance of ELN-XLN angular crossing in the asymptotic state for low spatial resolution models can be explained as follows. As shown by \cite{2022PhRvD.105j1301M}, the existence of ELN-XLN angular crossings becomes a necessary and sufficient condition for fast-flavor instability (FFI), but this condition holds only if all unstable modes including inhomogeneous modes are taken into account. We infer that FFCs in low resolution models undergo a non-linear saturation when lower $K$ modes become stable, yet the ELN-XLN angular crossing does not vanish entirely. To test this consideration, we carry out another linear stability analysis by using the result of the model with $N_z=10$. In this analysis, we compute the dispersion relation for the spatially averaged ELN-XLN angular distribution at the end of the simulation; the result is shown in Fig.~\ref{DR_novac}. As shown in the figure, modes around $K=0$ becomes stable, but there remain unstable modes at higher $|K|$, which is consistent with our argument.

We now turn our attention to the resolution dependence in the angular direction of neutrino momentum space. The results are summarized in Fig.~\ref{angdepe}. As shown in these plots, the results are not sensitive to the angular resolution. In fact, the model with $N_v=125$, corresponding to the lowest angular resolution in our models, is sufficient to accurately capture the asymptotic state of flavor conversions. The eigenvector for the unstable mode at $K=K_{\rm peak}$ provides insight into the feature, which is displayed as a blue line in Fig.~\ref{Qv_amp_hom}. As illustrated, the angular distribution of the eigenvector at $K=K_{\rm peak}$ does not exhibit a sharp profile, which can be resolved by $N_v=125$ grid points.

\begin{figure}
\begin{minipage}{0.5\textwidth}
    \includegraphics[width=\linewidth]{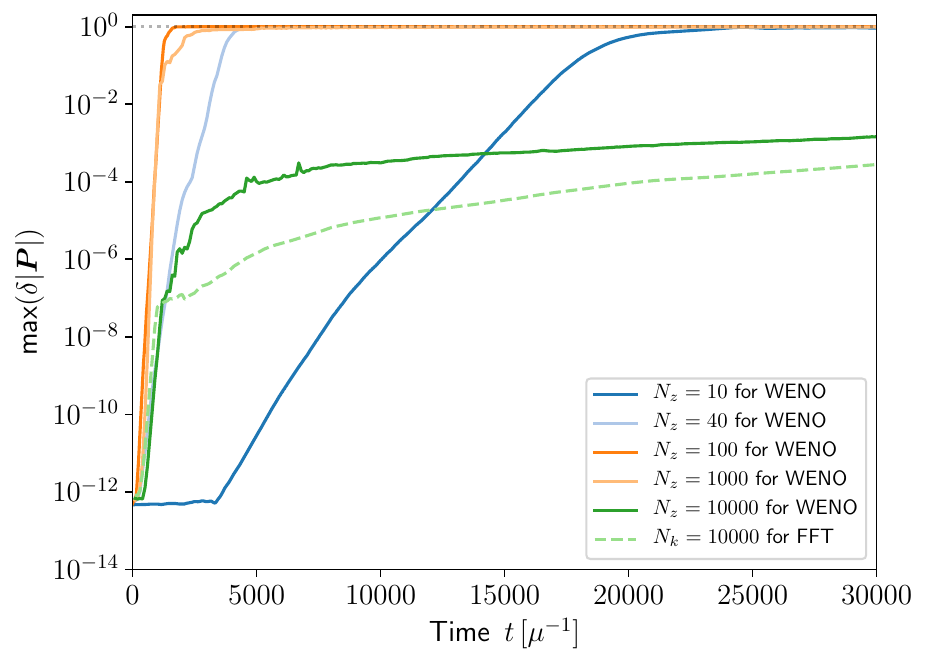}
\end{minipage}
    \caption{Maximum violation of $|\mathbf{P}|$, denoted as ${\rm max} (\delta |\mathbf{P}|)$, as a function of time. The color code is the same as Fig.~\ref{evo}. For comparison, we also display the result with pseudospectral method as a dashed line.
}
    \label{norm_novac}
\end{figure}

\begin{figure}
\begin{minipage}{0.5\textwidth}
    \includegraphics[width=\linewidth]{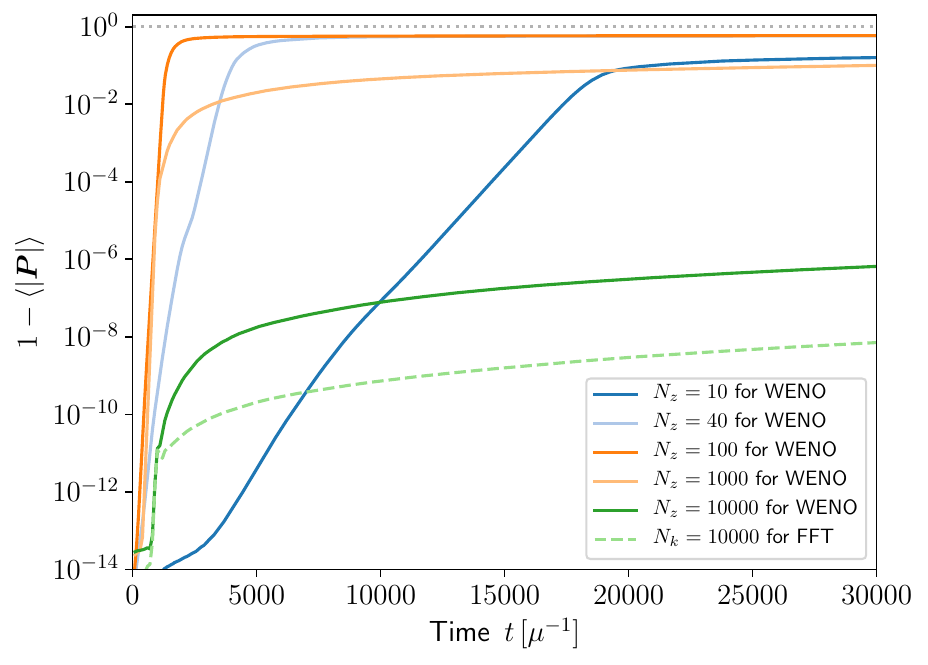}
\end{minipage}
    \caption{Same as Fig.~\ref{norm_novac} but we display the deviation from unity for the spatially averaged $|\mathbf{P}|$. This highlights the effect of artificial depolarization. See the text for more details.
}
    \label{norm_avg_novac}
\end{figure}

\subsection{Violation of $|\mathbf{P}|$ conservation}\label{subsec:vioP}

We also find another numerical problem in simulations with low spatial resolutions. As is well known, we can express the QKE with two flavor system in terms of three-dimensional polarization vectors of neutrinos ($\mathbf{P}$), which is related to the density matrix $\rho$ as $\rho = \rm{Tr} \rho/2 + \mathbf{P} \cdot \boldsymbol{\sigma}/2$, where $\boldsymbol{\sigma}$ represents the Pauli matrices. The norm of the polarization vector, $|\mathbf{P}|$, corresponds to a conserved quantity in flavor conversions, unless the collision term is included. In cases that $|\mathbf{P}|$ is distributed homogeneously in space at $t=0$, the neutrino advection should not alter the profile, meaning that $|\mathbf{P}|$ should be constant in time and space. However, the conservation is not guaranteed in most of the numerical schemes, because $|\mathbf{P}|$ is not solved directly as a primitive variable. For this reason, the conservation of $|\mathbf{P}|$ can serve as a useful metric to quantify the accuracy of simulations (see, e.g., Fig.~5 in \cite{2022PhRvD.106d3011R} and Fig.~9 in \cite{2020arXiv200500459B}). To evaluate this, we check the violation of $|\mathbf{P}|$ conservation in our models; the result is summarized in Fig.~\ref{norm_novac}. This figure is similar to Fig.~5 in \cite{2022PhRvD.106d3011R}, portraying the maximum violation of $|\mathbf{P}|$ from unity (the length of $|\mathbf{P}|$ is normalized by the true value). More specifically, we quantify the error of $|\mathbf{P}|$ as functions of $\boldsymbol{v}$ (or $\theta_{\nu}$) and $z$, and then plot the maximum value at each time snapshot. For comparison, we display a result obtained by the pseudospectral method in the same figure. As illustrated, the deviation is significant for low resolution simulations, and the overall trend is consistent with the resolution study in \cite{2020arXiv200500459B}. It should be mentioned that the slow increase of deviation in the model with $N_z=10$, corresponding to the model with the lowest spatial resolution, is simply due to a slow growth of flavor conversion (see also Fig.~\ref{evo}).

A few remarks should be made about the issue with $|\mathbf{P}|$. First, while the violation of the conservation is significant for $N_z=1000$ model, the numerical result of asymptotic state is reasonable (see Figs.~\ref{evo}~and~\ref{ang_dist}). This suggests that the violation does not have a direct impact on determining the asymptotic state. In fact, we can approximate the asymptotic state of FFC without the condition of $|\mathbf{P}|$ conservation (see \cite{2023PhRvD.107j3022Z,2023PhRvD.108f3003X}). Second, the violation accumulates with time even in the highest resolution model. This would be due to a turbulent-like property of flavor conversions in non-linear phase. As demonstrated in \cite{2020PhRvD.102j3017J,2021PhRvD.104h3035Z,2022PhRvD.106d3011R,2024PhRvD.109j3040U}, non-linear couplings of flavor waves generate small scale structures, which results in the energy cascade. This suggests that a certain amount of flavor energy leaks out from the maximum wave number which is determined by the grid width in finite volume methods. This trend can be seen in Fig.~\ref{norm_avg_novac}, in which we display the time evolution of the deviation of spatially averaged $|\mathbf{P}|$ from unity. As portrayed in the figure, the norm decreases with time even for the highest resolution simulation, indicating the occurrence of artificial depolarization. It should be mentioned that our simulations have much longer time span than those in \cite{2022PhRvD.106d3011R}, which allows us to highlight such an artificial depolarization. Our result suggests that many numerical simulations for quantum kinetic neutrino transport may suffer from such an artificial depolarization, which should be carefully monitored.

\subsection{Comparison with previous studies}\label{subsec:CompPrest}

The present study supports previous claims that high spatial resolutions are necessary for numerical modelings of FFCs (see, e.g., \cite{2021PhRvD.104j3023R,2021PhRvD.104j3003W,2022PhRvD.106d3011R,2022PhRvL.129z1101N}). On the other hand, there are disagreements between a previous study by \cite{2024arXiv240806422S} and ours. For instances, we find that both the growth rate of flavor coherency in the early phase and asymptotic states of flavor conversions strongly depend on the spatial resolution. However, the previous study of \cite{2024arXiv240806422S} showed that both of them are insensitive to the spatial resolution; in fact the overall time evolution of the system is essentially the same as that in a homogeneous case. As another qualitative discrepancy between the two studies is that the resolution dependence in the angular direction of neutrino momentum space is much milder than that found in \cite{2024arXiv240806422S}. These qualitative disagreements between the two studies lead to an opposite conclusion regarding requirements of spatial resolutions. To identify the source of the inconsistency, a detailed comparison study is required, but it is beyond the scope of this work. It should be emphasized, however, that the highest resolution models between our finite volume and pseudospectral methods are in good agreement with each other, which supports the reliability of our results.

Based on the present study, we finally provide an argument against a discussion regarding the length scale of neutrino flavor conversions in \cite{2024arXiv240806422S}. According to the previous paper, the time scale of flavor conversion is determined by $\mu^{-1}$ whereas the length scale is not. However, our present study shows that this discussion is misleading. As presented in Sec.~\ref{sec:linearsta}, the dispersion relation determines the characteristic scales of the system for both space and time in the early phase. For accurate modelings, it is essential to resolve all unstable modes; otherwise some special treatments (such as subgrid modeling) need to be implemented. 

It should also be mentioned that the neutrino flavor conversions in the non-linear phase have turbulent-like properties, in which small scale structures are developed due to non-linear mode couplings \cite{2020PhRvD.102j3017J,2021PhRvD.104h3035Z,2022PhRvD.106d3011R,2024PhRvD.109j3040U}. Incorrect modeling of the mode couplings can easily lead to numerical artifacts. This suggests that resolving the maximum unstable mode is just a minimal requirement and resolving higher wave number than $K_{\rm max}$, would be necessary for accurate modelings of neutrino flavor conversions in non-linear phase. In fact, the model with $N_z=1000$ has high enough resolution to capture the linear phase, but it still suffers from the large violation of the conservation of $|\mathbf{P}|$ (see Figs.~\ref{norm_novac}~and~\ref{norm_avg_novac}).

\section{Summary}\label{sec:summary}
Numerical simulations are the most useful tools to study non-linear physical processes. However, any numerical methods involve approximations of reality, indicating that the results must be carefully inspected to avoid drawing incorrect conclusions. In this paper, we address one of technical issues, resolution dependence, in numerical modelings of neutrino quantum kinetics, which has been much debate recently. We also provide a comprehensive analysis of the resolution dependence, which illustrates the mechanism of numerical artifacts generated by low resolutions.

We find that the growth rate of flavor conversion is underestimated, if the maximum growth mode is not properly captured. In these low resolution simulations, the asymptotic state of flavor conversions is qualitatively different from that in higher resolution models resolving all unstable modes of flavor instabilities. This is attributed to the fact that low spatial resolution models do not have abilities to erase ELN-XLN angular crossings which corresponds to a driving force of FFC. As a result, the angular resolution of neutrinos in low spatial resolutions becomes different from the highest resolution model. Such a numerical artifact can be understood by the dispersion relation computed based on the asymptotic state of spatially averaged neutrino angular distributions. This analysis reveals that growth rates for underresolved high wave numbers remain positive even in the asymptotic state, while only low $K$ modes (or resolved wave numbers) become stable. We also show that the conservation of $|\mathbf{P}|$ is significantly violated for low spatial resolution models. This is mainly due to artificial depolarization by leaking the flavor energy from the maximum wave number determined by spatial resolutions. We show that the artificial depolarization occurs even in the high resolution models, which potentially leads to numerical artifacts in long term simulations.

In conclusion, we emphasize that resolving all unstable inhomogeneous modes is a necessary condition for accurate modeling of neutrino quantum kinetics. This implies that unaffordable demands of computational resources to resolve the flavor waves make the global simulations intractable. We, hence, at the moment need to take effective approaches for global simulations such as the attenuation method or subgrid model. Since these approaches have been rapidly evolving during the last few years, more realistic numerical modelings of CCSN and BNSM under higher fidelity quantum kinetic neutrino transport are expected to be available in near future, which will bring us new insights into roles of neutrino flavor conversions on CCSN and BNSM.

\section{Acknowledgments}
The numerical simulations are carried out by using "Fugaku" and the high-performance computing resources of "Flow" at Nagoya University ICTS through the HPCI System Research Project (Project ID: 240079), XC50 of CfCA at the National Astronomical Observatory of Japan (NAOJ), and Yukawa-21 at Yukawa Institute for Theoretical Physics of Kyoto University. For providing high performance computing resources, Computing Research Center, KEK, and JLDG on SINET of NII are acknowledged. This work is supported by High Energy Accelerator Research Organization (KEK). H.N. is supported by Grant-inAid for Scientific Research (23K03468). M.Z. is supported by Grant-in-Aid for JSPS Fellows (Grant No. 22KJ2906) and JSPS KAKENHI Grant Number JP24H02245. L.J. is supported by a Feynman Fellowship through LANL LDRD project number 20230788PRD1.
\bibliography{bibfile}

\end{document}